# Barrier modification in sub-barrier fusion reactions using Dynamical cluster-decay and extended-Wong models

Raj Kumar, Manie Bansal and Raj K. Gupta

*Physics Department, Panjab University, Chandigarh-160014, India.*

***Abstract*— Fusion-evaporation cross-sections $\sigma_{evr}$ in reactions known for fusion hindrance phenomenon in coupled-channels calculations at below-barrier energies, are studied in terms of the dynamical cluster-decay model (DCM) of one of us (RKG) and Collaborators and the Wong formula extended by us, both based on proximity potential, by using the concept of "barrier modification" at sub-barrier energies, first advocated by Misicu and Esbensen for M3Y potential. The DCM is shown to contain the "barrier lowering" as its inbuilt characteristic, and the same is found essential, and introduced empirically, in the ($\ell$-summed) extended-Wong formula.**

***Keywords*— Fusion-hindrance, Barrier modification, Compound nucleus decay, Cluster-decay model, extended-Wong formula**

## I. Introduction

As a possible explanation to observed fusion hindrance phenomenon in coupled channel calculations (ccc) at extreme sub-barrier energies [1, 2], for fusion-evaporation cross-sections in reactions such as $^{58}$Ni+$^{58}$Ni, $^{64}$Ni+$^{64}$Ni and $^{64}$Ni+$^{100}$Mo, and capture cross-sections for $^{48}$Ca+$^{238}$U, $^{244}$Pu and $^{248}$Cm reactions, Misicu and Greiner [3] were the first to have shown that the M3Y-barriers, modified due to the *addition of repulsive core*, describe the capture cross-sections for $^{48}$Ca+$^{238}$U, $^{48}$Ca+$^{244}$Pu and $^{48}$Ca+$^{248}$Cm reactions, using equally well either the ccc or Wong formula [4], though for the case of different prescriptions for Q-values (Q-values used in experiments are not given in the published literature). Later [5], the same prescription for ccc was also found to be successful for fusion-evaporation cross-sections of all the three above mentioned $^{58,64}$Ni-based reactions. Note that the ccc could simply be sensitive to the so far unobserved, hence not-included, high-lying states. The repulsive core modifies [5] the shape of the inner part of the potential in terms of a thicker barrier (reduced curvature $\Delta\hbar\omega$) and shallower pocket.

The property of "lowering of barriers" at sub-barrier energies is also supported by the dynamical cluster-decay model (DCM) of pre-formed clusters by Gupta and Collaborators [6-8], where "barrier lowering" $\Delta V_B$ arises in a natural way in its fitting of the neck-length parameter $\Delta R$ (Fig. 1(a)). Very recently, we have also shown [9] that the Wong formula, extended to carry out its $\ell$-summation explicitly, also shows the necessity of "barrier modification" at sub-barrier energies, which can be affected empirically in terms of either the "barrier lowering" $\Delta V^{emp}$ (Fig. 1(b)) or "barrier narrowing" $\Delta\hbar\omega^{emp}$ via the curvature constant. In fact, the extended, $\ell$-summed Wong formula is a special case of DCM, more suitable for the capture or quasi-fission cross-sections where the incoming nuclei keep their identity.

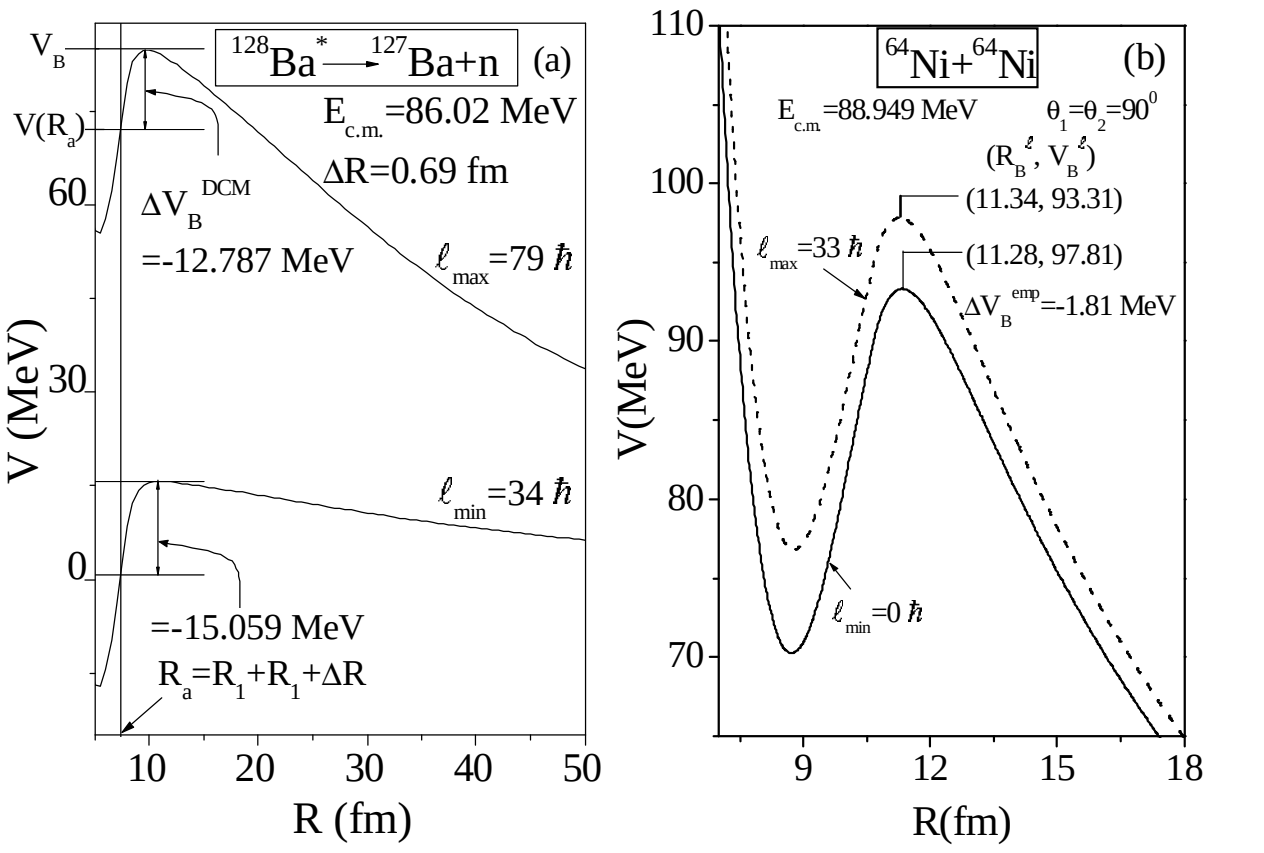

Fig. 1 (a) Definition of $\Delta V_B$, related to $\Delta R$, for LP=1n, in DCM. (b) The same for the incoming channel in $\ell$-summed Wong formula.

The dynamical cluster-decay model [8] for the decay of a hot and rotating compound nucleus (CN) formed in low-energy heavy ion reactions, is an alternate to the well known Hauser-Feshbach analysis and statistical fission models. The DCM, based on collective clusterization picture, considers the complete decay of a CN in to evaporation residues (evr), intermediate mass fragments (IMFs), fusion-fission (ff) and the competing quasi-fission (qf, or capture) processes as dynamical mass motions of preformed fragments or clusters through the interaction barrier, successfully applied to evr cross-sections $\sigma_{evr}$ in $^{64}$Ni+$^{100}$Mo→$^{164}$Yb$^*$, IMFs emission in $^{48}$Cr$^*$, $^{56}$Ni$^*$ and $^{116}$Ba$^*$ decays, IMFs and ff in decays of $^{116,118,122}$Ba$^*$, ff (and qf) of $^{246}$Bk$^*$ formed in $^{11}$B+$^{235}$U and $^{14}$N+$^{232}$Th and the three processes of evr, ff, and the competing qf (equivalently, capture) in $^{48}$Ca+$^{238}$U, $^{244}$Pu, and $^{154}$Sm reactions. DCM has the in-built property of "lowering of barriers" at sub-barrier energies through its single neck-length parameter $\Delta R$. The nuclear proximity potential used in DCM is, so far, of Blocki *et al.* [10] and the one obtained recently [11] for the Skyrme nucleus-nucleus interaction in the semiclassical extended Thomas Fermi (ETF) approach.

For the capture (or quasi-fission) process, since the two incoming nuclei do not loose their identity, the preformation factor $P_0^\ell$ =1, and the DCM expression for cross-section reduces to that of Wong model. Thus, whereas the capture process is treated on similar footings in both the Wong and DCM ($P_0^{\ell}=1$), the fusion-evaporation cross-sections in Wong need the “barrier modification” and the same in DCM arises because $P_0^{\ell}>0$.

A point of difference in the two models (Wong and DCM) is that the penetrability $P_0^{\ell}$ in Wong formula is calculated in the Hill-Wheeler [12] approximation of inverted harmonic oscillator for the interaction potential $V_{\ell}(R)$ calculated for the *incoming channel*, whereas the same in DCM is the WKB integral whose first turning point $R_a$ (given by Eq. [5]) is defined through a neck-length parameter $\Delta R$ for the best fit to, say, the data on fusion-evaporation cross-section or fission, which also contains the “barrier lowering” effects in it *for each decay channel* [6,7]. In this contribution, we present the results obtained for the use of DCM and extended-Wong model for the illustrative reaction $^{64}Ni+^{64}Ni$, whose first brief report was made earlier in [7].

## II. Methodolgy

*Dynamical Cluster-decay Model (DCM) and Wong model:*

The DCM [8] is based on the collective coordinates of mass (and charge) asymmetry $\eta$ (and $\eta_Z$) [$\eta=(A_1-A_2)/(A_1+A_2)$; $\eta_Z=(Z_1-Z_2)/(Z_1+Z_2)$], and relative separation R. For the de-coupled $\eta$-, R-motions, in terms of the $\ell$-partial waves, the DCM defines the fragment formation or compound-nucleus decay cross section for oriented nuclei as [6-8]

$$\sigma(E_{c.m.},\theta_i)=\frac{\pi}{k^2}\sum_{\ell=0}^{\ell_{max}}(2\ell+1)P_0P_{\ell};\qquad k=\sqrt{\frac{2\mu E_{c.m.}}{\hbar^2}} \quad (1)$$

where, $P_0$, the pre-formation probability, refers to $\eta$-motion and $P_{\ell}$, the penetrability, to R-motion, both depending on $\ell$, T, $\beta_{\lambda i}$ and $\theta_i$.

The $P_0$ is given by the solution of stationary Schrödinger equation in $\eta$, at a fixed $R=R_a$, the first turning point(s) of the penetration path(s) shown in Fig. 1(a) for different $\ell$-values,

$$\left\{-\frac{\hbar^2}{2\sqrt{B_{\eta\eta}}}\frac{\partial}{\partial\eta}\frac{1}{\sqrt{B_{\eta\eta}}}\frac{\partial}{\partial\eta}+V(R,\eta,T)\right\}\psi^{\nu}(\eta)=E^{\nu}\psi^{\nu}(\eta) \quad (2)$$

with $\nu$=0,1,2,3...., referring to ground-state ($\nu$ =0) and excited-states solutions. Then, the probability

$$P_0(A_i)=|\psi_R(\eta(A_i))|^2\sqrt{B_{\eta\eta}}\frac{2}{A}, \quad (3)$$

where, for a Boltzmann-like function,

$$|\psi|^2=\sum_{\nu=0}^{\infty}|\psi^{\nu}|^2\exp(-E^{\nu}/T) \quad (4)$$

For the first turning point $R_a$, in the case of the decay of a hot CN, we use the postulate [6-8],

$$\begin{aligned}R_a&=R_1(\alpha_1,T)+R_2(\alpha_2,T)+\Delta R(\eta,T)\\&=R_t(\alpha,\eta,T)+\Delta R(\eta,T)\end{aligned} \quad (5)$$

with radius vectors

$$R_i(\alpha_i,T)=R_{0i}(T)\left[1+\sum_{\lambda}\beta_{\lambda i}Y_{\lambda}^{(0)}(\alpha_i)\right] \quad (6)$$

and the temperature dependent nuclear radii $R_{0i}(T)$ for the equivalent spherical nuclei,

$$R_{0i}=[1.28A_i^{1/3}-0.76+0.8A_i^{-1/3}](1+0.0007T^2) \quad (7)$$

Here, for fixed $R=R_a$, the fragmentation potential

$$\begin{aligned}V_R(\eta,T)=&-\sum_{i=1}^{2}[B_i(A_i,Z_i,T)]+V_P(R,A_i,\beta_{\lambda i},\theta_i,T)\\&+V_C(R,Z_i,\beta_{\lambda i},\theta_i,T)+V_{\ell}(R,A_i,\beta_{\lambda i},\theta_i,T),\end{aligned} \quad (8)$$

Thus, the DCM treat all decay processes as the dynamical collective mass motion of pre-formed clusters through the interaction barrier, i.e., use the actual decay channels like the light-particles (LPs) for $\sigma_{evr}$, etc.

Wong formula [4] is Eq. (1) of DCM with $P_0$=1 (denoted, $\ell$-summed Wong model), with P calculated in Hill-Wheeler approximation as

$$P_{\ell}(E_{c.m.},\theta_i)=\left[1+\exp\left(\frac{2\pi(V_B^{\ell}(E_{c.m.},\theta_i)-E_{c.m.})}{\hbar\omega_{\ell}(E_{c.m.},\theta_i)}\right)\right]^{-1} \quad (9)$$

Instead of solving this equation explicitly, Wong carried out the $\ell$-summation approximately, and obtained $\theta_i$-integrated $\sigma$ in terms of the $\ell$=0 based barrier for the incoming channel, thereby ignoring the “barrier modification” due to its $\ell$-dependence. In $\ell$-summed Wong model, we carry out the $\ell$-summation explicitly for a best fit to the data [9].

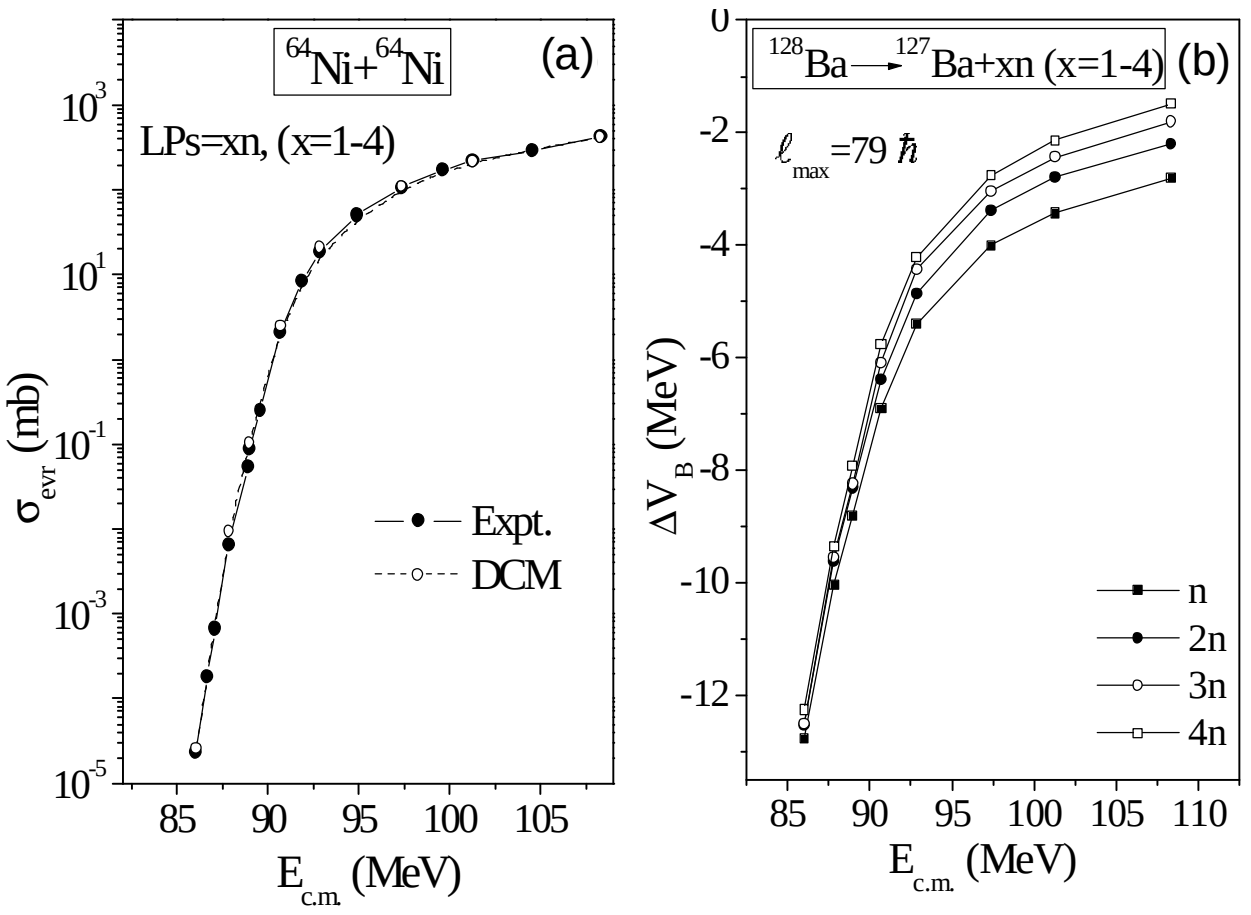

Fig. 2 In DCM, (a) fitted ER cross-sections for LPs=1n-4n, and (b) “Barrier lowering” $\Delta V_B$ at $\ell_{max}$ for the best fitted $\sigma_{evr}$.

## III. CALCULATIONS

Fig. 2 shows the results of our calculation, using the DCM, for light-particles constituting the evr. For the best fit of neck-length parameter $\Delta R$ to experimental $\sigma_{evr}$ in Fig. 2(a), the deduced $\Delta V_B$ at $\ell_{max}$ in Fig. 2(b) show that “barrier lowering” at sub-barrier energies is clearly present in DCM. Similarly, the results of $\ell$-summed Wong formula in Fig. 3(a) clearly show the necessity of lowering the barrier of the incoming channel in Fig. 3(b) at below-barrier energies.

## IV. Conclusions

Barrier lowering at sub-barrier energies in $^{64}Ni+^{64}Ni$ reaction is an essential property contained in the DCM, and is required empirically for the $\ell$-summed Wong formula.

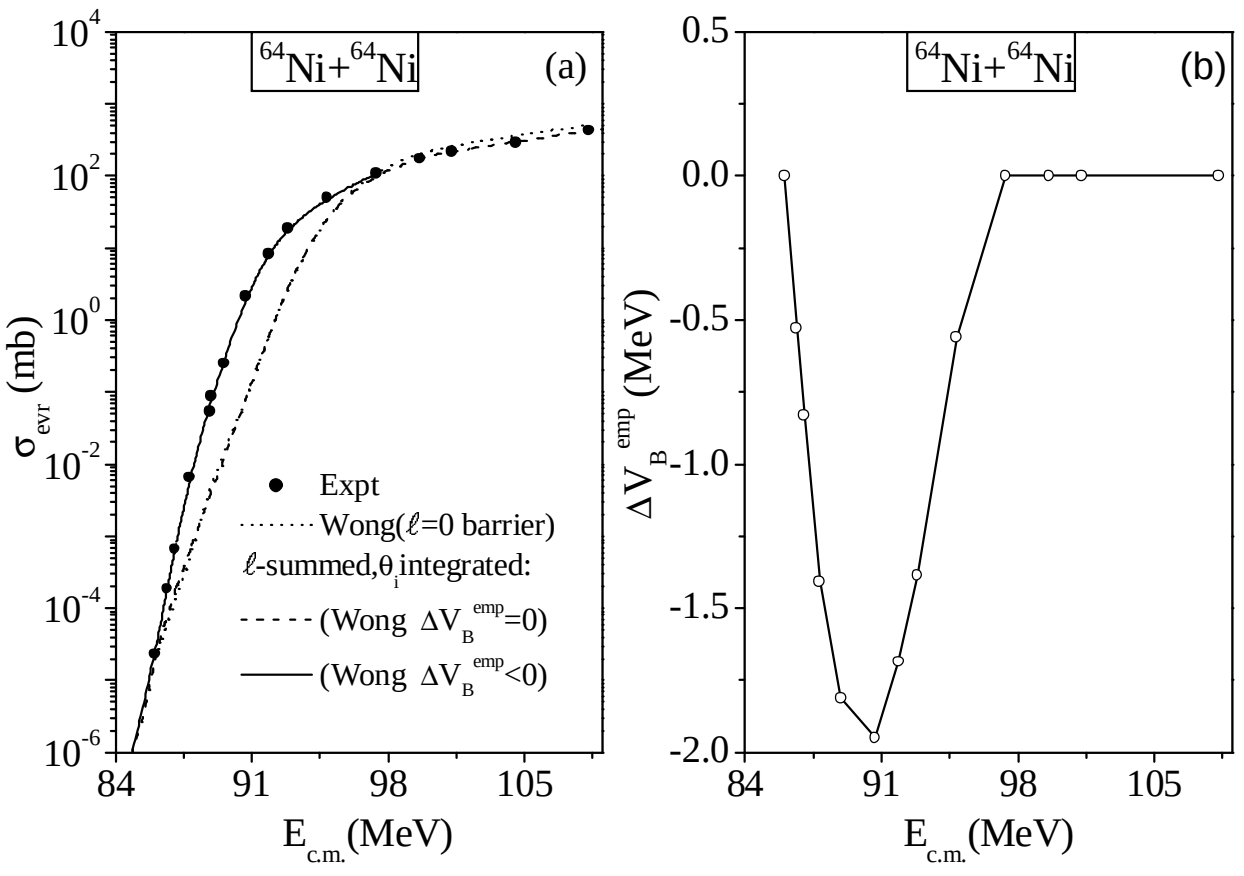

Fig. 3 Same as for Fig. 2, but for the incoming channel in ℓ-summed Wong formula.

## ACKNOWLEDGMENT

The financial support from the Department of Science and Technology (DST), Government of India, and the Council of Scientific and Industrial Research (CSIR), New Delhi, is gratefully acknowledged.